## ZTF J172132.75+445851.0: A Possible New Cataclysmic Variable of the VY Sculptoris Type

Klaus Bernhard[1], Christopher Lloyd[2], Peter Frank[3], Wolfgang Moschner[4]

1) Bundesdeutsche Arbeitsgemeinschaft für Veränderliche Sterne e.V., Germany
American Association of Variable Star Observers (AAVSO), USA
email: klaus.bernhard@liwest.at , ORCID 0000-0002-0568-0020
2) Department of Physics and Astronomy, University of Sussex, Brighton BN19QH, UK
email: C.Lloyd@sussex.ac.uk , ORCID 0000-0003-4323-1960
3) Bundesdeutsche Arbeitsgemeinschaft für Veränderliche Sterne e.V., Germany
email: frank.velden@t-online.de
4) Bundesdeutsche Arbeitsgemeinschaft für Veränderliche Sterne e.V., Germany
email: wolfgang.moschner@gmx.de

**Abstract:** We report the discovery of ZTF J172132.75+445851.0 as a new possible VY Sculptoris variable, with an orbital period of 0.109765426(44) days. Survey observations from 2005 to the present reveal significant activity in the system, with brightness variations ranging between 18.1 and 21.3 (zr) magnitudes, including deep fades exceeding an amplitude of 3 magnitudes.

### 1   Introduction

Cataclysmic variables (CVs) are close binary systems, typically consisting of a white dwarf accreting matter from a low-mass secondary companion that fills its Roche lobe, see e.g., [1], [2]. Among the various types of CVs, the VY Sculptoris stars (GCVS-type NL) are nova-like variables characterized by irregular fading events caused by temporary reductions in mass transfer rates, differing from typical nova outbursts observed in other CVs. As a result, they are sometimes called "anti-dwarf novae". These decreases in brightness are believed to result from a temporary reduction or cessation of mass transfer, although the mechanism is unclear.

VY Sculptoris stars exhibit two distinct components of variability: a short-period, orbital component, typically with a period of 0.10-0.17 days, and a long-term component characterized by irregular variability in combination with rare pronounced fading events. This makes the light curve of VY Scl stars very characteristic, making confusion with other types of variable stars unlikely. One such object was discovered 10 years ago by one of us (K.B.) and published as 1SWASP J022916.91-395901.4 [3]. As of December 2024, a total of 185 VY Scl stars (type NL/VY) are included in the International Variable Star Index (VSX) of the AAVSO [4], and for comparison there are a similar number of UGSS stars, and about five times as many UGSU stars.

## 2 Observation and data analysis

We discovered ZTF J172132.75+445851.0 during a systematic visual inspection of objects with very short periods (<0.2 days) in the Zwicky Transient Facility (ZTF) catalog of variable stars. There it is listed as an EA variable with the unusually short period of 0.10976 d [5]. This catalog includes approximately 780,000 periodic variables with classifications.

The catalog relies on data obtained from the Zwicky Transient Facility (ZTF), a time-domain survey operational since 2017 at Palomar Observatory. The ZTF camera, featuring e2v CCD231-C6 devices, is mounted on the Palomar 48-inch Samuel Oschin Schmidt Telescope. Covering three passbands (g, r, and partially i), it achieves a limiting magnitude of 20.5 mag, making ZTF data highly suitable for variable-star investigations [6], [7], [8]. Further data were acquired through the 'cyan' c band (420–650 nm), and the 'orange' o band (560–820 nm) of the ATLAS Forced Photometry web service [9], [10], [11]. The c and o bands overlap, and are centred to the red of the ZTF g and r bands. Data were also taken from the Catalina Sky Survey [12] covering the years 2005 – 2013 in the V band.

In 2024, one of the authors (Wolfgang Moschner) conducted supplementary observations using a private ASA Astrograf at Nerpio/Spain.

**Observations**
Telescope: 400 mm ASA Astrograph f/3.7 (f = 1471 mm)
Camera: FLI 16803 CCD Camera
Filter: CV
Exposure Time: 300 seconds
Data analysis:   Phoranso 1.0.4.8 (https://www.cbabelgium.com/Phoranso/)
StarCurve (L. Pagel, https://www.bav-astro.eu/index.php/weiterbildung/tutorials)
The weighted mean of five comparison stars was used for the differential photometry. Flatfield correction was not employed.

The essential data of ZTF J172132.75+445851.0 are listed below. The positions (Epoch J2000) are computed by VizieR and are not part of the original Gaia DR3 data.

| | | |
|---|---|---|
| Identifiers | **ZTF J172132.75+445851.0** | |
| | Cross-IDs: | |
| | Gaia DR3 1361411552301991040 | |
| | GALEX J172132.7+445850 | |
| | TIC 1400461048 | |
| | ATO J260.3865+44.9808 | |
| ZTF | Magnitude Range (r) | 18.1-21.3 mag |
| Gaia DR3 [13] | Right Ascension (J2000) | 17h21m32.76s |
| | Declination (J2000) | +44°58'51.45" |
| | G-band mean magnitude (350-1000 nm) 18.55 mag | |
| | Integrated BP mean magnitude (330-680 nm) 18.77 mag | |
| | Integrated RP mean magnitude (640-1000 nm) 17.65 mag | |
| | BP-RP   1.1182 mag | |
| | Gmag(abs) | 8.11 mag |
| | Plx | 1.0207±0.0996 mas |
| | Distance | 1239.6 pc |

## 3 Results

**Long period variability**

The ZTF (zr, zg) and the ATLAS (o, c) light curve of ZTF J172132.75+445851.0, is shown in Figure 1. The system is very active with variability on various time scales: There are annual trends, where it can change by a magnitude or more, and within each season short-term variations can generate larger scatter, most notably in 2020. Equally, there are periods of relative stability, and in this time frame the 2018 season shows the least variation. Furthermore, it has periods of short, deep, rapid fades as in 2020 and 2024, and a longer fade in 2019. Compared with other well-known VY Scl systems like MV Lyr and TT Ari [1], this system is very active and does not show long periods at a bright, constant magnitude.

Despite the trends and other variations there is a consistent colour difference when the system is bright, which is best determined in the 2018 season giving <g> = 18.95, <r> = 18.38, and <g-r> = 0.57. When the system is faint it is also bluer, and during the deep fade of 2020 <g-r> = 0.10. The bluer state during deep fades could indicate an increased dominance of the white dwarf or the inner hot accretion disk. These changes are clearly visible in Figure 2, which shows detail of the light curve from 2018 to early 2020.

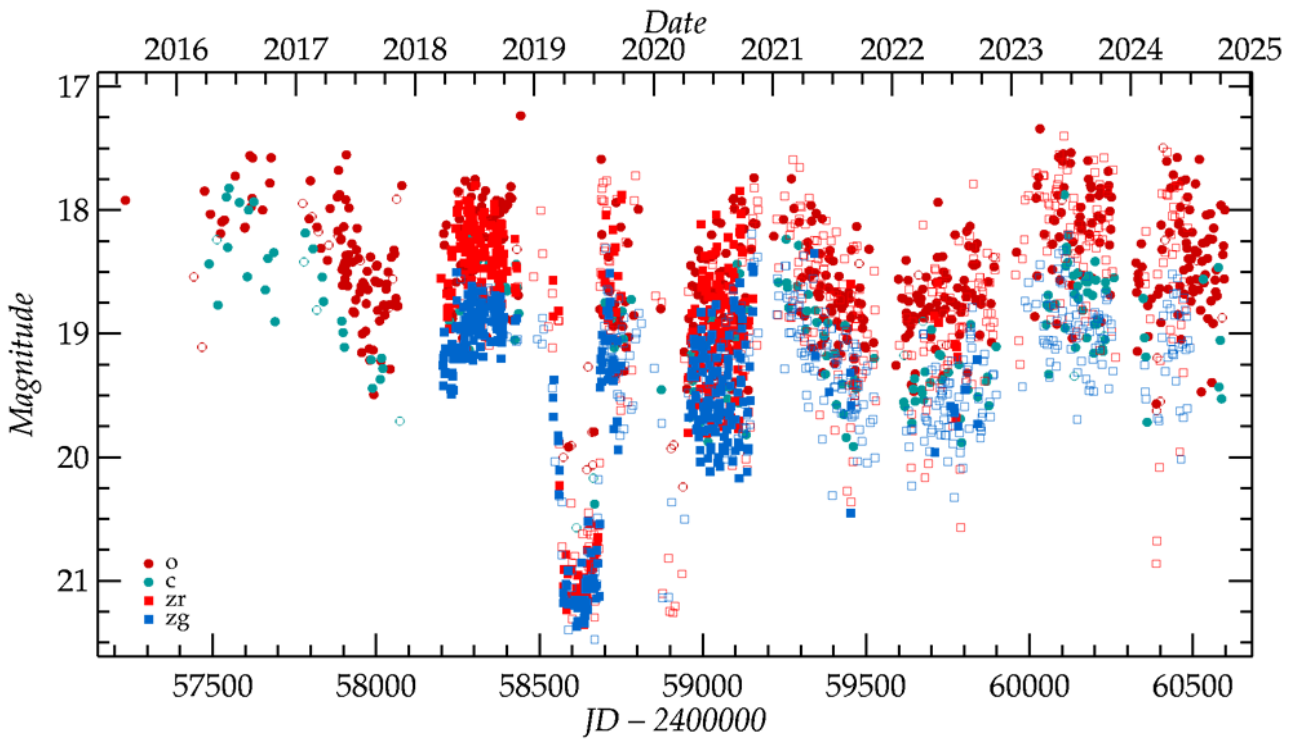

Figure 1: Light curve of the ATLAS o and c data from 2015 to the present and ZTF r and g data from 2018. The filled symbols are daily means of typically 5 observations, while the open symbols of the respective colours are individual measurements. Only the positive ATLAS observations are shown for clarity.

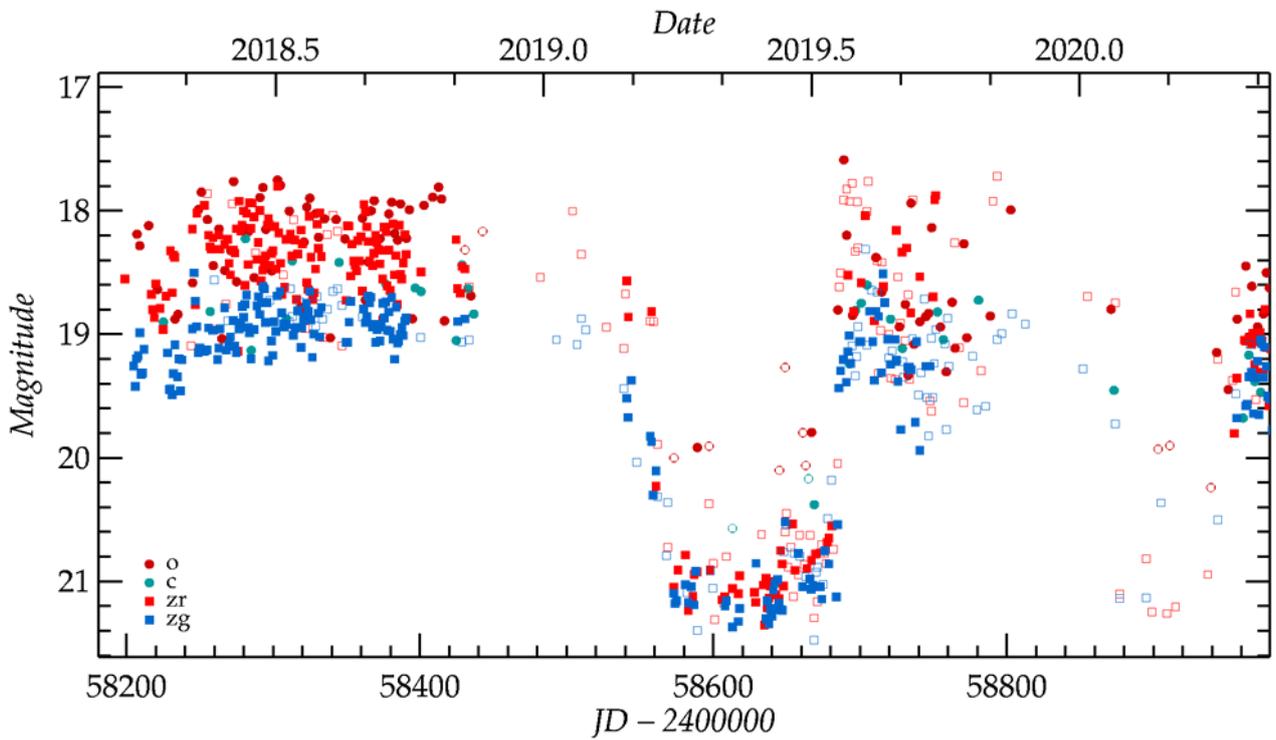

Figure 2: Detail of the ZTF r and g and ATLAS o and c light curve showing the most stable observing season of 2018 followed by the deep fading event in 2019, and the less well-observed fade of 2020. The change in the g-r colour between the bright and faint phases is clear.

Very similar behaviour can be seen in the long-term V-band light curve of the CSS data (2005 – 2013) with annual variations of up to two magnitudes and periods of short, deep, rapid fades in 2007 and 2008 (see Figure 3).

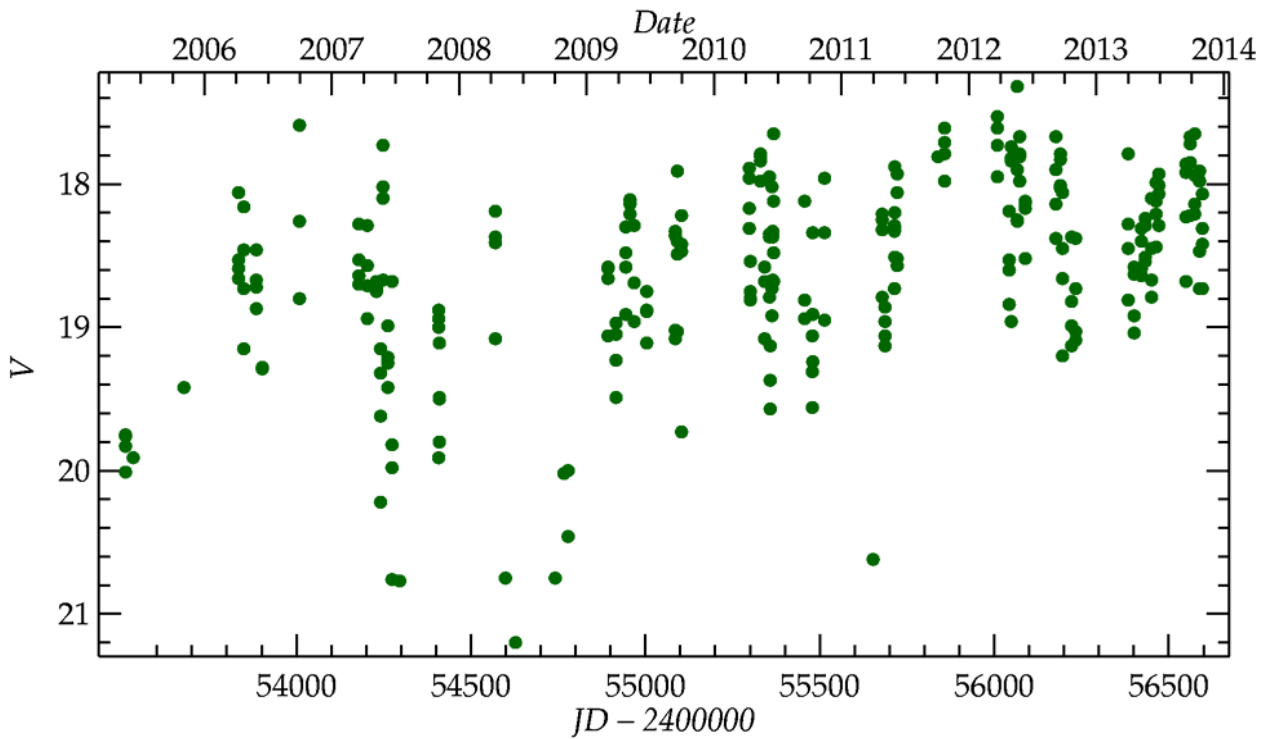

Figure 3: Light curve from CSS V band

**Short-period variability**

The large overall "scatter" in the g and r light curves of Figure 1 is at least partly due to periodic short-term variability. It is best seen in 2018 when the mean brightness is relatively constant and the dispersion is low. An analysis of a restricted subset of the 2018 ZTF data (JD = 2458240 – 2458440, excluding the early faint points) using the Anova method of Peranso [14], which is recommended for non-sinusoidal periodic variations and implements the algorithm described by Schwarzenberg-Czerny, identified a period of 0.109766(1) d. The individual bands were fitted with a 4-harmonic Fourier series giving P = 0.1097664(34) and 0.1097674(35) d, in the g and r bands respectively. The period from i-band data is consistent with these, but the uncertainty is an order of magnitude larger. The phased light curve of this subset is presented in Figure 4, folded on the ephemeris of the r-band data, and also shows the individual fits using the r-band period.

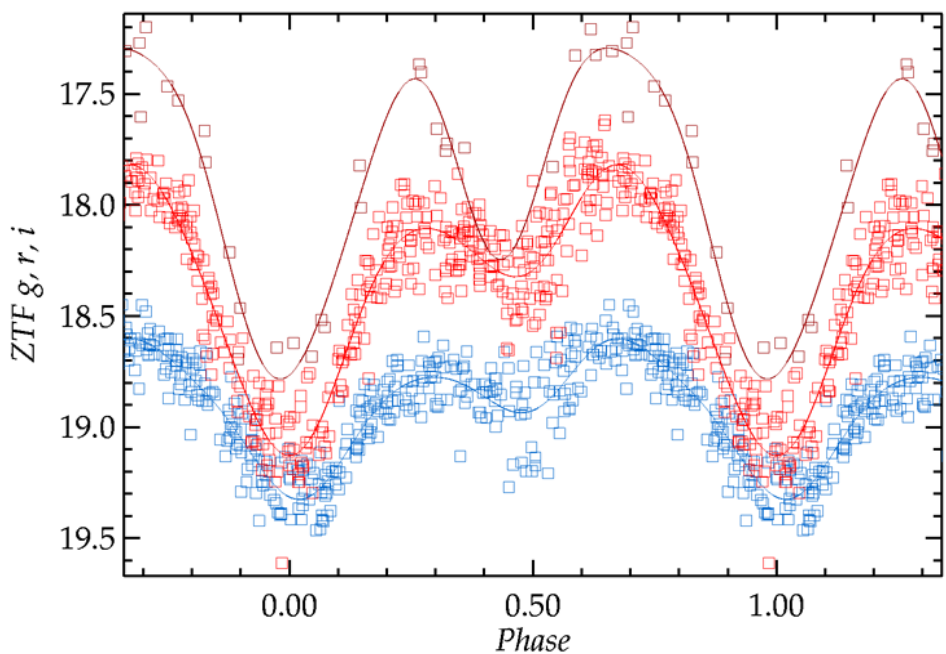

Figure 4: Phased light curve of the restricted subset of ZTF g (blue squares), r (red squares) and i (reddish-brown squares) band data (JD = 2458240 – 2458440) with P = 0.1097674 d. The lines show the fits to the individual bands with that period, and the differences are shown in Figure 5.

ZTF J172132.75+445851.0 shows its strongest variability in the i-band, as evidenced by the phased light curve, followed by the r-band, with the least variability in the g-band. The system appears bluest at the primary minimum, with colour indices of g−r~0.2 and r−i~0.3. Notably, g−r undergoes only slight changes between phases 0.25 and 0.75, while r−i reaches a minimum at both 'eclipses'. The changes in colour are shown in Figure 5 as the differences between the fits in Figure 4.

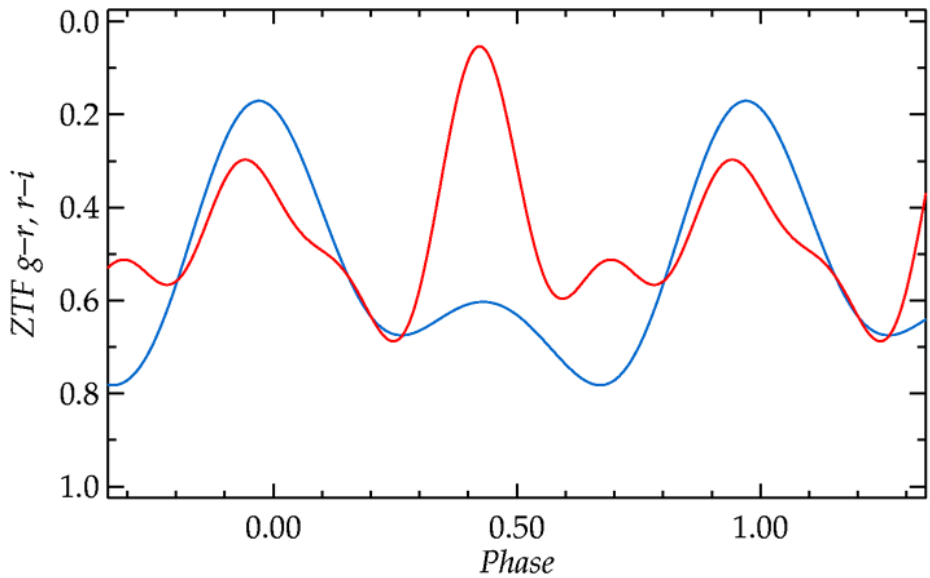

Figure 5: Phased ZTF g-r (blue) and r-i (red) band light curves from the fits to the restricted subset of data (JD = 2458240 – 2458440) with the period 0.1097674 d in Figure 4.

The short-period variability is detectable to some extent in the ZTF and ATLAS data in most years, despite the longer-term changes in brightness, but it cannot be detected in 2019 during the prolonged deep fade. Also, the short-period variability can be detected in the latter half of the CSS data.

Finally, the short-period variability was also confirmed in our own observations from Nerpio/Spain in August and September 2024 (Figure 6). The observations were made on two nights, separated by nine days, and there is a significant change in level between them. These were also fitted with a 4-harmonic Fourier series, but with a free offset between them. The best period is 0.109712(26) d, which is consistent with the better determined ZTF period at about the 2-sigma level.

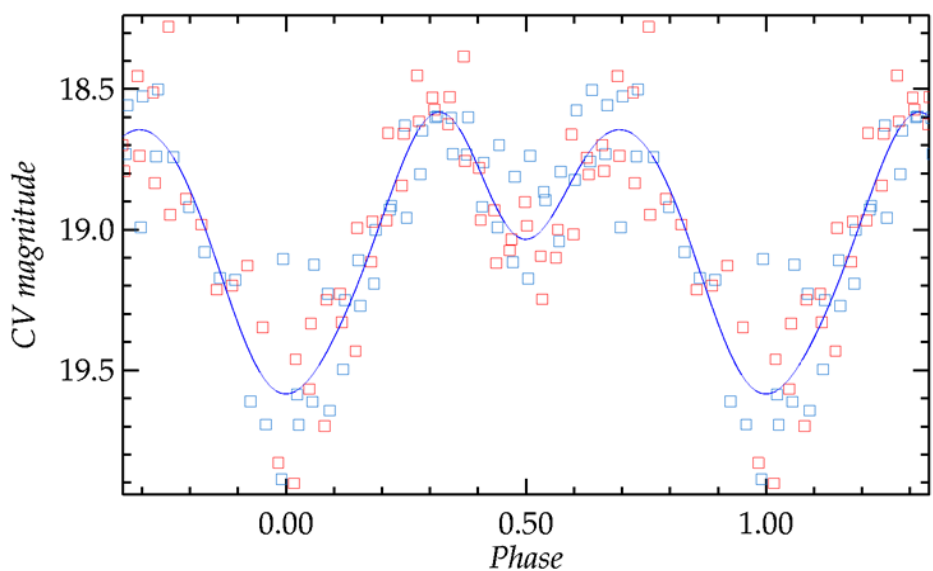

Figure 6: Phased g (blue squares), r (red squares) light curves with the period 0.109712 d of observations from Nerpio/Spain.

The visibility of the orbital period in most of the data provides an opportunity to firstly, determine its stability and confirm that it is an orbital variation, and secondly, to improve the period. The period is only weakly visible in the CSS data from 2009 on, and the time of minimum is only realistically measurable for the years 2011 – 2013. For the ZTF and ATLAS data the times of minima were measured for each season with any linear trends removed. Most of the reliable minima come from the ZTF r and i bands, and the ATLAS o band, partly due to the greater coverage, and the larger amplitude. The data sets were fitted with a 4-harmonic Fourier series using a fixed period. In most cases the secondary minima are poorly constrained, and those that are used tend to have large uncertainties. A weighted fit to the timings gives an ephemeris of

$$HJD_{Min\ I} = 2458319.7837(4) + 0.109765426(44) \times E$$

and the corresponding O–C diagram is shown in Figure 7, where the different bands are identified as in the earlier figures. Secondary minimum is assumed to occur at phase 0.5, and there is no evidence that it deviates systematically from that. Despite the seasonal trends and more rapid changes, the orbital signature remains a significant feature of the variations, except during the deep minima.

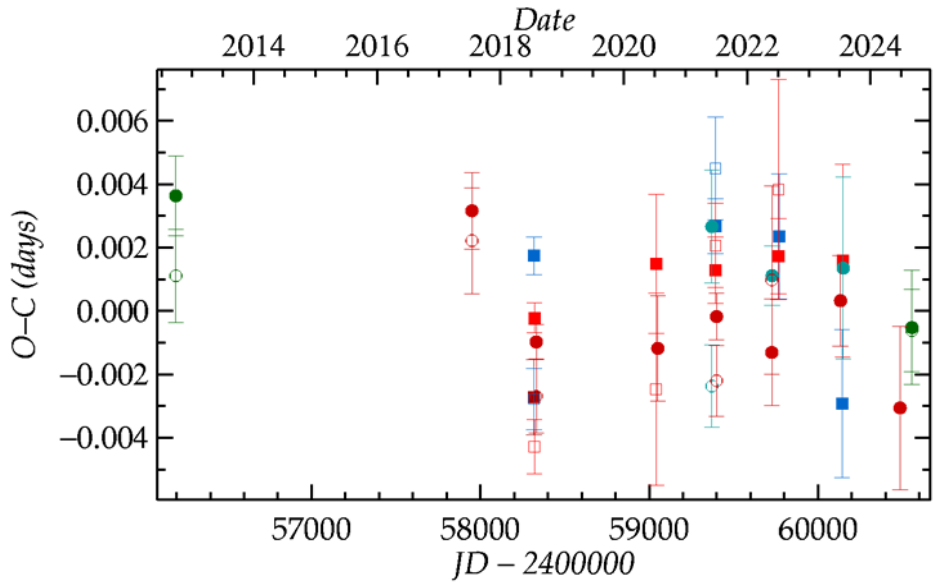

Figure 7. The O–C diagram of the times of minima constructed using the ephemeris, showing the different bands.

## 4  Discussion

The available survey data and own observations of ZTF J172132.75+445851.0 provide evidence for the coexistence of long-term variability, such as deep fading events indicative of reduced or halted mass transfer, with short-period variability at 0.1097654 days, which corresponds to the orbital period of the system. These characteristics align with the expected behaviour of accretion-dominated VY Scl systems in which orbital modulations and variations in accretion disk dynamics are observed. The faint absolute magnitude of $M_G \sim 8$ derived by Gaia DR3 is also consistent with the typical luminosity range of cataclysmic variables, further supporting the classification as VY Scl system. Compared with other VY Scl systems ZTF J172132.75+445851.0 is very active, and does not remain at one level from one season to the next, like e.g., MV Lyr, TT Ari, or MP Gem, but appears to be more similar to V794 Aql or KR Aur [1],[15]. Its colour variation with brightness is also similar to V794 Aql. A significant part of the variation is orbital, over a magnitude in r, so this may indicate a high inclination near 90°, which may not be the case for other systems.

Table 1. Times of minima

| HJD | Error | Min | Cycle no. | O – C | Band | Source |
|---|---|---|---|---|---|---|
| 2456195.7165 | 0.0013 | 1 | -19351.0 | 0.0036 | V | CSS |
| 2456195.7689 | 0.0015 | 2 | -19350.5 | 0.0010 | V | CSS |
| 2457949.8774 | 0.0012 | 1 | -3370.0 | 0.0032 | o | ATLAS |
| 2457949.9313 | 0.0017 | 2 | -3369.5 | 0.0022 | o | ATLAS |
| 2458317.8097 | 0.0006 | 1 | -18.0 | 0.0018 | zg | ZTF |
| 2458317.8601 | 0.0010 | 2 | -17.5 | -0.0027 | zg | ZTF |
| 2458319.7835 | 0.0005 | 1 | 0.0 | -0.0002 | zr | ZTF |
| 2458319.8344 | 0.0009 | 2 | 0.5 | -0.0042 | zr | ZTF |
| 2458320.7688 | 0.0012 | 1 | 9.0 | -0.0028 | zi | ZTF |
| 2458330.8691 | 0.0005 | 1 | 101.0 | -0.0009 | o | ATLAS |
| 2458330.9223 | 0.0012 | 2 | 101.5 | -0.0026 | o | ATLAS |
| 2459040.8344 | 0.0022 | 1 | 6569.0 | 0.0016 | zr | ZTF |
| 2459040.8853 | 0.0030 | 2 | 6569.5 | -0.0024 | zr | ZTF |
| 2459047.7482 | 0.0008 | 1 | 6632.0 | 0.0002 | zi | ZTF |
| 2459047.8008 | 0.0026 | 2 | 6632.5 | -0.0021 | zi | ZTF |
| 2459050.8204 | 0.0017 | 1 | 6660.0 | -0.0011 | o | ATLAS |
| 2459371.0100 | 0.0018 | 1 | 9577.0 | 0.0028 | c | ATLAS |
| 2459371.0598 | 0.0013 | 2 | 9577.5 | -0.0023 | c | ATLAS |
| 2459377.8122 | 0.0015 | 1 | 9639.0 | -0.0005 | zi | ZTF |
| 2459377.8639 | 0.0013 | 2 | 9639.5 | -0.0036 | zi | ZTF |
| 2459392.7435 | 0.0009 | 1 | 9775.0 | 0.0028 | zg | ZTF |
| 2459392.8002 | 0.0016 | 2 | 9775.5 | 0.0046 | zg | ZTF |
| 2459392.8519 | 0.0010 | 1 | 9776.0 | 0.0014 | zr | ZTF |
| 2459392.9076 | 0.0013 | 2 | 9776.5 | 0.0022 | zr | ZTF |
| 2459398.8875 | 0.0007 | 1 | 9831.0 | -0.0001 | o | ATLAS |
| 2459398.9404 | 0.0011 | 2 | 9831.5 | -0.0021 | o | ATLAS |
| 2459725.9874 | 0.0017 | 1 | 12811.0 | -0.0012 | o | ATLAS |
| 2459726.0446 | 0.0030 | 2 | 12811.5 | 0.0011 | o | ATLAS |
| 2459727.9656 | 0.0009 | 1 | 12829.0 | 0.0013 | c | ATLAS |
| 2459765.8353 | 0.0012 | 1 | 13174.0 | 0.0019 | zr | ZTF |
| 2459765.8923 | 0.0035 | 2 | 13174.5 | 0.0040 | zr | ZTF |
| 2459769.8972 | 0.0020 | 1 | 13211.0 | 0.0025 | zg | ZTF |
| 2460132.8895 | 0.0014 | 1 | 16518.0 | 0.0005 | o | ATLAS |
| 2460141.7773 | 0.0023 | 1 | 16599.0 | -0.0027 | zg | ZTF |
| 2460145.7333 | 0.0030 | 1 | 16635.0 | 0.0018 | zr | ZTF |
| 2460148.9163 | 0.0029 | 1 | 16664.0 | 0.0015 | c | ATLAS |
| 2460485.8918 | 0.0026 | 1 | 19734.0 | -0.0029 | o | ATLAS |
| 2460555.9246 | 0.0018 | 1 | 20372.0 | -0.0003 | CV | Moschner |
| 2460555.9794 | 0.0013 | 2 | 20372.5 | -0.0004 | CV | Moschner |

The system also lies 29 arcsec (about 2-sigma) from the unidentified X-ray source 1RXS J172133.5+445919 listed in the ROSAT All-Sky Survey Faint Source Catalog [16]. The hardness ratios are HR1=0.20±0.25 and HR2=0.67±0.23, and are more consistent with chromospheric activity than a white dwarf.

Only one object of this type (V0434 Gem) is listed with a shorter period in the AAVSO-VSX catalog as of December 2024. This makes ZTF J172132.75+445851.0 a particularly intriguing candidate for further investigation. We encourage additional spectroscopic and photometric observations to confirm the classification and gain deeper insights into its physical properties.

**Acknowledgements**

This research has utilized the SIMBAD/VIZIER database and Aladin, operated at CDS, Strasbourg, France, the International Variable Star Index (VSX) database, operated at AAVSO, Cambridge, Massachusetts, USA, the NASA/IPAC Infrared Science Archive and the SAO/NASA Astrophysics Data System, USA. The authors thank Udo Reffke (BAV) for helpful comments.